\documentstyle[aps,prb,twocolumn,epsf,eqsecnum]{revtex}

\begin{document}

\twocolumn[\hsize\textwidth\columnwidth\hsize\csname @twocolumnfalse\endcsname

\draft

\title{The effect of phase fluctuations on the single-particle properties of 
the underdoped cuprates}

\author{Hyok-Jon Kwon\cite{emailjon} and Alan T. Dorsey\cite{emailalan}}
\address{Department of Physics, University of Florida, Gainesville,
FL 32611-8440}

\date{September 16, 1998}

\maketitle

\begin{abstract}

We study the effect of order parameter 
phase fluctuations on the single-particle properties of fermions in the 
underdoped cuprate superconductors using a phenomenological low-energy 
theory.  We identify the fermion-phase field coupling as the Doppler-shift
of the quasiparticle spectrum induced by the fluctuating superfluid velocity 
and we calculate the effect of these fluctuations on the fermion self-energy. 
We show that the vortex pair unbinding near the superconducting transition  causes a
significant broadening in the fermion spectral function, producing a 
pseudogap-like feature.  We also discuss  the specific heat and show that
the phase fluctuation effect is visible due to the short coherence length.   

\end{abstract}


\vskip2pc]

\section{Introduction}

One of the more intriguing properties of the cuprate superconductors is the 
existence of a ``pseudogap'' regime in the normal phase, 
which develops when these materials
are underdoped and at temperatures below a characteristic temperature
$T^*$ (see Fig.~1). 
Numerous experiments, including NMR,\cite{nmr} 
neutron scattering,\cite{neutron} 
infrared and optical conductivity,\cite{optical} transport,\cite{transport}
Raman scattering,\cite{raman} specific heat,\cite{specific-heat}
angle-resolved photoemission (ARPES),\cite{arpes} and scanning
tunneling spectroscopy (STS),\cite{sts} indicate that in this 
regime there is a 
gap in the excitation spectrum, while transport measurements show
that the sample is normal and that the superfluid density 
is zero.\cite{randeria97} 
The ARPES and STS experiments on Bi$_2$Sr$_2$CaCu$_2$O$_{8+\delta}$
have established that the pseudogap
evolves smoothly from the superconducting gap. More significantly, the 
ARPES results show that the pseudogap in the normal phase possesses the
same $d_{x^2-y^2}$ symmetry as the superconducting gap; the node along the 
$(\pi/2,\pi/2)$ direction in the superconducting phase evolves into 
an extended gapless region with increasing temperature in the pseudogap 
regime, while the gap maximum in the $(\pi,0)$ direction has only a very 
weak temperature dependence. Taken collectively, these results
demonstrate that the pseudogap in the normal phase 
is a remnant of the quasiparticle gap in the superconducting phase.

The theoretical challenge in understanding the pseudogap regime is then to 
reconcile the existence of a quasiparticle gap with 
the absence of superconducting order.
The usual strategy in studying superconductivity is to start from a 
normal phase consisting of well-defined Landau quasiparticles and 
to then ask what interactions lead to a superconducting instability. 
Here, the experiments suggest that we start from the
{\it superconducting} phase---skirting the issue of the origin of the 
pairing---and ask what mechanisms would destroy phase coherence 
while preserving the gap in the excitation spectrum.  

\begin{figure}[ht]
\epsfxsize=6.0cm \epsfbox{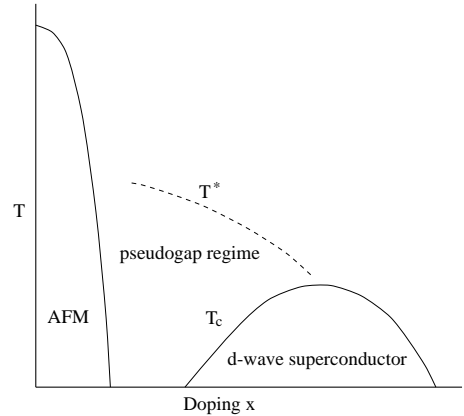}
\bigskip
\caption{Schematic phase diagram of a cuprate superconductor as a function
of temperature $T$ and doping $x$. AFM denotes the antiferromagnetic insulating
phase.}
\label{phase-diagram}
\end{figure}

The mechanism which we explore in this paper,  originally 
suggested by Emery and Kivelson,\cite{Emery} is that 
the local superconducting order parameter
$\Delta({\bf x})  = \Delta_0 e^{i\theta ({\bf x})}$ has a well defined
amplitude $\Delta_0$ (set by energies of order $T^*)$ but a 
fluctuating phase $\theta $, such that 
$\langle \Delta({\bf x})\rangle 
= \Delta_0 \langle e^{i\theta ({\bf x})}\rangle = 0$ 
in the pseudogap regime.  The temperature scale for phase 
fluctuations $T^{\rm phase}$ is controlled by the zero-temperature 
superfluid stiffness $n_s(0)/4m$; the actual critical temperature 
$T_c={\rm min}\{ T^{\rm phase},T^{\rm MF}\}$, with $T^{\rm MF}$ the 
mean-field transition temperature predicted in BCS theory. 
Emery and Kivelson argued that in conventional
low-$T_c$ superconductors, which have a large superfluid stiffness, 
$T^{\rm phase}\gg T_c$, so that phase fluctuations are irrelevant and 
the transition is BCS-like. 
However, cuprate superconductors are doped Mott insulators
which have a small superfluid stiffness (of order the doping $x$), 
so that $T^{\rm phase} \sim T_c$ and the transition is controlled
by phase ordering.  This naturally explains the Uemura 
scaling,\cite{Uemura} $T_c\sim n_s(0)/m$, observed in the cuprate
superconductors.   In addition, the reduction of the phase-stiffness 
$n_s(T)/4m $ is approximately
linear in temperature due to the $d$-wave symmetry, almost all the way to
$T_c$,\cite{Jacobs} in contrast to conventional $s$-wave 
superconductors, where $n_s(T)/4m $ rapidly drops to zero
near $T_c$.\cite{Fetter} Therefore the temperature range for which $n_s(T)/4m $
is a small number is much wider than in the conventional superconductors, 
and the effect of phase fluctuations can be observed even 
far from $T_c$. 

In this paper we investigate the effects   of phase fluctuations
on the fermion single-particle properties and the thermodynamics near the 
superconducting transition, 
without regard to the details of a 
microscopic model. In addition to the smallness of $n_s(T)/4m $,
we assume that the superconducting
gap and the normal state pseudogap have a $d$-wave symmetry and that the
system is two-dimensional due to the weakness of the interlayer coupling. 
We find that phase fluctuations, and in particular transverse phase
fluctuations, can have a pronounced effect on the spectral function, 
density of states, and specific heat near the transition. 

A brief overview of the paper is as follows:
In Sec.~II  we construct a low-energy effective theory of fermion
quasiparticles coupled to the classical phase fluctuations by 
integrating out the fast momentum degrees of freedom.
In Sec.~III we obtain the self-energy of the (neutral) fermion
quasiparticles and calculate the single-particle density of states
and the spectral broadening.
We observe that the peak in the spectral function 
and the density of states near the gap is 
significantly broadened by the phase fluctuations (vortex pair
unbinding) near the
transition temperature, exhibiting a pseudogap-like behavior. 
In Sec.~IV we consider the effect of the phase fluctuations on the
specific heat near the transition. We find that the 
vortex-antivortex unbinding transition causes a
visible peak in the specific heat just above the transition temperature
due to the short coherence length of the materials.
Appendix A  is a compilation of some useful results on the 
Berezinskii-Kosterlitz-Thouless transition, and Appendix B contains some
technical details on the calculation of the fermion 
self-energy. 

\section{Effective low-energy theory}
\label{s:eff}
In this section we derive a low-energy effective theory of fermion
quasiparticles coupled to the phase fluctuations of the superconducting
order parameter, assuming that the amplitude fluctuations are negligible
near $T_c$.
For simplicity, we consider an $s$-wave symmetry of the gap in this
section.
Extensions to other gap symmetries are straightforward.

We begin with the mean-field BCS model whose partition function is
given by
\begin{equation}
{\cal Z} = \int {\cal D}c_{\sigma} {\cal D}c^{\dag}_{\sigma} 
 {\cal D}\Delta {\cal D}\Delta^* \, e^{-S},
\end{equation}
where
\begin{eqnarray}
S &=& \int d^2x\,\int_0^\beta d\tau \left[\sum_\sigma
c^{\dag}_\sigma({\bf x},\tau) 
\left(\partial_\tau -{\nabla^2 \over 2m_0} -\mu\right)
c_\sigma({\bf x},\tau)\right.  \nonumber \\
& & \ \ +\left. \Delta ({\bf x},\tau)c^{\dag}_\uparrow({\bf x},\tau)
  c^{\dag}_\downarrow({\bf x},\tau) + {\rm h.c.} 
 +{1\over g}|\Delta({\bf x},\tau)|^2\right], 
\label{Act1}
\end{eqnarray}
with the interaction strength $g>0$.
The pairing field $\Delta ({\bf x},\tau)=|\Delta({\bf x},\tau)|e^{i\theta ({\bf x},\tau)}$;
in what follows we will assume that the amplitude of the order parameter 
is constant, $|\Delta({\bf x},\tau)|=\Delta$, and concentrate on the 
phase degree of freedom, $\theta $. Also we have not considered 
the effect of Coulomb interactions.
In order to couple the fermions
to the $\theta  $ field more explicitly, we perform a gauge transformation
$ \psi_\sigma({\bf x},\tau) = c_\sigma({\bf x},\tau)  
e^{-i\theta ({\bf x},\tau)/2}$, with $\psi_\sigma$ the field operators for
neutral fermions.\cite{mpaf} Then we obtain a new action for the
neutral fermions  as 
\begin{eqnarray}
S &=& \int d^2x\, \int_0^\beta d\tau \Bigg\{ \sum_\sigma
\psi^{\dag}_\sigma({\bf x},\tau) 
\Big[    (\partial_\tau+ i\partial_\tau \theta/2)\nonumber \\ 
& & \quad - {(\nabla +i\nabla \theta  /2)^2 \over 2m_0}
-\mu \Big] \psi_\sigma({\bf x},\tau)  \nonumber \\
& & \quad + \Delta \psi^{\dag}_\uparrow({\bf x},\tau)
  \psi^{\dag}_\downarrow({\bf x},\tau) + {\rm h.c.} 
 + {1\over g}\Delta^2\Bigg\}. 
\label{Act2}
\end{eqnarray} 
The last term in Eq.~(\ref{Act2}) is a constant and will be dropped in 
what follows.
\begin{figure}[t]
\center
\epsfxsize=5.0cm \epsfbox{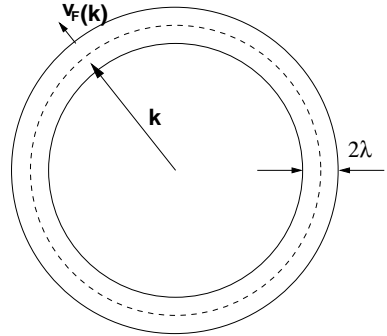}
\bigskip
\caption{Hilbert space of the low-energy effective theory.
After the inner and outer side of the thin momentum shell degree of freedom
has been integrated out, we are left with a theory of fermion 
quasiparticles living in a thin momentum shell of thickness $ 2\lambda $.
The dashed line denotes the Fermi surface.
The direction of the Fermi velocity ${\bf v}_F({\bf k})$ is in parallel
with the momentum ${\bf k}$ of the fermion quasiparticle $ \psi ({\bf k})$. }
\label{f:FS}
\end{figure}

As it stands, Eq.~(\ref{Act2}) is hard to study, so
we integrate out the fast momentum degrees of freedom and focus on the
low-energy properties of the system.\cite{andreev}
Namely, $\psi \rightarrow \psi +\psi^{\prime} $ where
$\psi^{\prime}({\bf x}) = 
\sum_{||{\bf k}|-k_F|>\lambda} \psi_{\bf k} e^{i{\bf k\cdot x}}$ is the
fast momentum degree of freedom which is to be integrated out of the
partition function and $\psi({\bf x}) = 
\sum_{||{\bf k}|-k_F|<\lambda} \psi_{\bf k} e^{i{\bf k\cdot x}}$ is the 
low-energy quasiparticle.
Then 
\begin{eqnarray}
{\cal Z}&=&\int {\cal D}\psi^{\prime} {\cal D}\psi^{\prime \dag}
{\cal D}\psi {\cal D}\psi^{\dag} {\cal D}\theta  \, e^{-S}  \nonumber \\
&=&{\cal N}\int {\cal D}\psi {\cal D}\psi^{\dag} {\cal D}\theta  
\, e^{-S_{\rm eff}},
\end{eqnarray}
and we are left with the momentum degrees of freedom in the vicinity
of the Fermi surface if we take $\lambda \ll k_F $. Since we will be considering
the effect of fluctuating vortices on the fermion self-energy, and the
characteristic vortex core size is of order the coherence length $\xi_0$, 
we will take $\lambda$ of order $1/\xi_0$.  We are therefore working in the
limit $k_F \xi_0\gg 1$; in the cuprates it is estimated that 
$k_F\xi_0 \sim 10$,\cite{coherence} so our approximation is appropriate 
for these materials.

The effective action of the fermions near the Fermi surface and the phase 
field is
\begin{eqnarray} 
S_{\rm eff} &= & \sum_\sigma\Big{[} T\sum_{{\bf k},\omega}
\psi^{\dag}_\sigma ({\bf k},\omega )
 \psi_\sigma ({\bf k},\omega )
 ( i\omega- k^2/2m +\mu ) \nonumber \\
&&~ + \int d^2x \int_0^\beta d\tau\, \psi^{\dag}_\sigma
({\bf x},\tau)(i\partial_\tau \theta /2 \nonumber \\
&&~-{i\over 2m} \nabla \theta \cdot \nabla
) \psi_\sigma({\bf x},\tau)\Big{]}
  \nonumber \\
&& +{1\over 2}T\sum_{{\bf q},\nu}{n_f \over 4m} {\bf q}^2
 \theta({\bf q}, \nu)\, \theta({\bf -q}, -\nu)
\nonumber \\
&&~+T\sum_{{\bf k},\omega}\Delta \psi^{\dag}_\uparrow ({\bf k},\omega )
\psi^{\dag}_\downarrow ({\bf -k},-\omega ) +{\rm h.c.}.
\end{eqnarray}
Here we have neglected higher orders in $ q/k \ll 1$ 
since $k\approx k_F$ and $|{\bf q}| < \lambda$, and we
have taken $c^{\dag}_\sigma c_\sigma \approx n_f $, the total number of fermions.
Since $\bf k$ is close to the Fermi surface, we may linearize the
fermionic spectrum as $k^2/2m-\mu \equiv \xi_{\bf k}
\approx v_F  (|{\bf k}|-k_F) $. 
The resulting effective action is expressed in terms of the Nambu 
spinor notation,
$\hat{\psi} = (\psi_{\uparrow}, \psi^{\dag}_{\downarrow}) $, as
\begin{equation}
S_{\rm eff}= S_0 + S_I,
\end{equation}
with 
\begin{eqnarray}
S_0&=& T\sum_{{\bf k},\omega} \hat{\psi}^{\dag} \hat{G}_0^{-1} \hat{\psi} 
     \\ \nonumber
 & & \quad +{1\over 2}\int_0^\beta d\tau 
\int d^2x \, {n_f \over 4m} \,[\nabla \theta ({\bf x},\tau )]^2
\label{Act0}
\end{eqnarray}
and
\begin{eqnarray}
S_I & =&\int d^2x \int_0^\beta d\tau \hat{\psi}^{\dag}
({\bf x},\tau)\Bigl(i\partial_\tau \theta /2 \\ \nonumber
 & & \quad  -{i\over 2m} \nabla \theta \cdot \nabla
       \Bigr) \hat{\psi}({\bf x},\tau) .
\end{eqnarray}
Here
\begin{equation}
 \hat{G}_0^{-1} = \left( 
\begin{array}{cc} 
i\omega - \xi_{\bf k} & -\Delta \\
-\Delta & i\omega + \xi_{\bf k} \\
\end{array}
 \right),
\end{equation}
with $\hat{G}_0$ the bare Green's function for the neutral fermions,
\begin{equation}
 \hat{G}_0 = \left( 
\begin{array}{cc} 
{\cal G}_0({\bf k},\omega) &  {\cal F}_0({\bf k},\omega) \\
{\cal F}_0({\bf k},\omega) & -{\cal G}_0({\bf k},-\omega)\\
\end{array} 
\right),
\end{equation}
\begin{eqnarray}
{\cal G}_0({\bf k},\omega) & =&{i\omega + \xi_{\bf k} \over 
(i\omega - \xi_{\bf k}) (i\omega +\xi_{\bf k})-\Delta^2},\nonumber\\
 {\cal F}_0({\bf k},\omega)&=&{ \Delta \over (i\omega - \xi_{\bf k})
               (i\omega +\xi_{\bf k})-\Delta^2}.
\end{eqnarray}

The microscopic physics relevant to the length scale $L < \xi_0 $
(the vortex core structure, for instance) has been integrated out to 
renormalize such quantities as the effective mass $m$ and the
residual interactions between the quasiparticles which we have not
included here. Therefore, in this effective theory  the vortices are 
treated as point-like entities.

In studying the finite temperature superconductor to normal metal
transition, it should be sufficient to consider only the static
fluctuations of the phase, and we may suppress the time-dependence
in $\theta $ and retain only the spatial fluctuation. The interaction
term is then reduced to
\begin{equation}
S_{\rm I} = - T\sum_{{\bf k, q},\omega} m{\bf v}_F({\bf k})\cdot {\bf v}_s ({\bf
 q})
\hat{\psi}^{\dag}({\bf k+q},\omega ) \hat{\psi}({\bf k},\omega ).
\label{doppler}
\end{equation}
Here we take
$ {\bf k}/m \approx {\bf v}_F({\bf k})$
where ${\bf v}_F({\bf k})$ is the Fermi velocity at the Fermi surface point
closest to the wave vector $\bf k$ (see Fig.~\ref{f:FS}),
and ${\bf v}_s ({\bf q}) = \int d^2x\, e^{-i{\bf q\cdot x}}\nabla
\theta({\bf x}) /2m $ is the
superfluid velocity. One should bear in mind that ${\bf v}_s({\bf q})$
includes both longitudinal and transverse components. 

When the interaction is expressed in form of Eq.~(\ref{doppler}) 
we see that $S_{\rm I}$ is the Doppler shift of the fermion spectrum
due to a non-zero superfluid velocity.  This becomes more explicit if we
consider the $q\rightarrow 0$ limit of $S_I$, neglecting the ${\bf q}$ dependence
of the field operators, so that the superfluid velocity can be subsumed into 
the fermion Green's function as
\begin{equation}
 \hat{G}^{-1} = \left(
\begin{array}{cc}
i\omega - \xi_{\bf k} - {\bf p}_F\cdot{\bf v}_s & -\Delta \\
-\Delta & i\omega + \xi_{\bf k}  - {\bf p}_F\cdot{\bf v}_s\\
\end{array}
 \right),
\end{equation}
where ${\bf p}_F \equiv m {\bf v}_F ({\bf k})$. 
A semiclassical approximation of this form has recently been used
by Franz and Millis\cite{franz98} to study phase fluctuations in the 
underdoped cuprates.  They incorporated the phase fluctuations by averaging the 
Green's function over a Gaussian distribution of velocity fluctuations. 
Our approach is rather different---we treat the coupling to the 
fluctuating phase using a self-consistent perturbation theory (see Sec.~III)
in order to calculate the fermion self-energy due to phase 
fluctuations.  

Finally, we can check if the effective theory correctly gives the
effective action of the phase fluctuations\cite{rama} by integrating
out the remaining fermionic degrees of freedom.
By expanding $e^{-S_{I}}$ to second order in $\theta $, and
integrating out the fermion fields, we obtain
\begin{eqnarray}
S[\theta  ]& \approx &
{1\over 2}\sum_{{\bf k,q},\omega}\left(  v_F^2  
q^2 \over 4 \right) {\rm Tr} \left[ \hat{G_0}({\bf k},\omega ) 
\hat{G_0}({\bf k+q},\omega ) \right]\nonumber \\ 
&&\times\theta({\bf q})\theta({\bf -q})
 +{1\over 2T} \sum_{\bf q} {n_f \over 4m}  q^2 
\, \theta({\bf q})\theta({\bf -q}).
\label{PhAct}
\end{eqnarray}
The superfluid density at temperature $T$ is\cite{Fetter}
\begin{eqnarray}
n_s(T) &=&
n_f + \lim_{{\bf q} \rightarrow 0} \lim_{\nu \rightarrow 0}
{T\over m}\sum_{ {\bf k}, \omega} ({\bf k+q}/2)_i ({\bf k+q}/2)_i\nonumber \\
&& ~\times {\rm Tr} \left[ \hat{G_0}({\bf k},\omega ) 
\hat{G_0}({\bf k+q},\omega +\nu ) \right].
\end{eqnarray}
Therefore, for $|{\bf q}| \ll \Delta /v_F$, we recognize that 
Eq.~(\ref{PhAct}) has the following form:
\begin{equation}
S[\theta  ]\approx {1\over 2T} \sum_{\bf q} 
{n_s(T) \over 4m} q^2 \theta({\bf q})\theta({\bf -q}).
\label{ph-eff}
\end{equation}
We see that the stiffness of the phase fluctuations can be
identified as the superfluid density at $T$.
This result holds even when we consider
unconventional order parameter symmetry.

Equation~(\ref{ph-eff}) can serve as a convenient effective theory to describe
the critical properties of the Berezinskii-Kosterlitz-Thouless (BKT) 
transition.\cite{bkt}
The macroscopic measurable quantities are often insensitive to the
detailed microscopic fermion theory except through the parameter
$n_s(T)/m$, and the effective description of phase fluctuations is enough
to characterize the transitions in macroscopic quantities.
A brief discussion of the BKT transition and derivation of the effective
classical action are included in Appendix A for completeness. 
We should stress that we are {\it not} proposing that the 
superconducting transition in the cuprates is of the BKT type; 
weak interplanar coupling will produce a transition in the 
$d=3$ $XY$ universality class, sufficiently close the transition.
However, outside this regime two-dimensional vortex fluctuations 
will dominate the low-energy physics,  and we expect the physics 
discussed in the subsequent sections to pertain. Indeed, recent time-domain
terahertz spectroscopy measurements\cite{corson98} of the complex conductivity 
$\sigma(\omega)$ of underdoped 
${\rm Bi}_2{\rm Sr}_2{\rm CaCu}_2{\rm O}_{8+\delta}$,  which provide a direct 
probe of the superfluid density, show that the dynamics is well described
by the BKT picture, lending credence to the model developed in this paper.

\section{Single-particle properties of fermions with phase
fluctuations}

The underdoped cuprate superconductors can be regarded as doped Mott
insulators with a small charge carrier density (superfluid density),
in other words, $n_s(0)/m \sim x E_F $ where $x$ is the doping concentration.
To calculate the fermionic self-energy, we have to perform a perturbative
expansion in $\theta  $ from Eq.~(\ref{Act2}),
but close to $T_c$ and for a small 
superfluid density, there is no obvious small parameter. We need a way to
select a set of meaningful diagrams. 
The saving grace in this case is that the volume (perimeter) of the 
Fermi surface is large, so that we can select the diagrams of
leading order in $1/N(0)$. One way of keeping track of the diagrams is 
to introduce $n$ fermion species and perform a $1/n $ expansion and
at the end take $n \rightarrow 1 $. This amounts to summing over ring
diagrams. Other diagrams are smaller by factors of $\lambda / k_F$
since only the ring diagrams involve summations over the Fermi
surface perimeter in this low-energy effective theory.
In this section we study the effect of phase fluctuations on the
single-particle properties of the fermions using this machinery,
both at $T<T_{BKT}$ and $T>T_{BKT}$.
For the rest of the paper, we reinstate the angular dependence of
the gap function $\Delta (\phi)$, and perform calculations assuming
a $d_{x^2-y^2}$-wave symmetry, $\Delta (\phi) = \Delta_0 \cos 2\phi$,
neglecting the temperature dependence in $\Delta_0$ since it 
changes smoothly through the transition in the underdoped cuprates.

\begin{figure}[t]
\epsfxsize=8.0cm \epsfbox{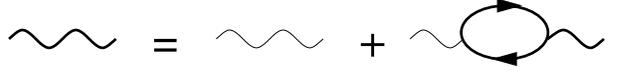}
\bigskip
\caption{The propagator of the velocity field in the leading order in
$\lambda / k_F$. The thin (thick) wiggly line is the
bare (full) propagator. The ring in the figure
is the polarization of the fermions.}
\label{f:ring}
\end{figure}
\begin{figure}[t]
\epsfxsize=8.0cm \epsfbox{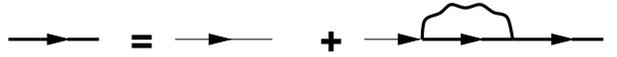}
\bigskip
\caption{The Dyson equation of fermion Green's function. The
wiggly line is the velocity propagator. The thin (thick) solid line 
is the bare (full) Green's function.}
\label{f:dyson}
\end{figure}

We first calculate the velocity-velocity correlation function from the effective 
theory we obtained in Sec.~\ref{s:eff}  by summing over
the ring diagrams as in Fig.~\ref{f:ring}.  
We find that
\begin{equation}
\langle v_s^{\alpha}({\bf q}) v_s^{\beta}(-{\bf q}) \rangle_{\rm ring} 
= {\langle v_s^{\alpha}({\bf q}) v_s^{\beta}(-{\bf q}) \rangle_0\over
1+ \langle v_s^{\alpha}({\bf q}) v_s^{\beta}(-{\bf q}) \rangle_0
    \Pi_{\alpha\beta}({\bf q},\omega)}~,
\end{equation}
where
\begin{eqnarray}
\Pi_{\alpha\beta}({\bf q},\omega) &=& \sum_{\bf k}
\left[m{\bf v}_{F}({\bf k})\right]_{\alpha} \left[m{\bf v}_{F}({\bf k})\right]_{\beta}
\\ \nonumber 
& & \quad \times {\rm Tr}\, [\hat{G}_0({\bf k+q},\omega ) \hat{G}_0({\bf k},\omega )].
\end{eqnarray}
Now we calculate the fermion self-energy correction from the Dyson
equation, 
\begin{eqnarray}
\hat{\Sigma}({\bf k}, \omega ) 
&=&  \sum_{\bf q} \left[m{\bf v}_{F}({\bf k})\right]_{\alpha} 
\left[m{\bf v}_{F}({\bf k})\right]_{\beta}\\  \nonumber 
& & \quad \times \langle v_s^{\alpha}({\bf q}) v_s^{\beta}(-{\bf q}) \rangle_{\rm ring}
\hat{ G}({\bf k-q}, 
\omega ) ~,
\label{dyson}
\end{eqnarray}
where $\hat{ G} $ is the full fermion Green's function, given 
self-consistently by $\hat{ G} ^{-1} =\hat{ G}_0 ^{-1} - \hat{\Sigma}$.
As shown in Appendix~A, the velocity-velocity correlation function 
can be resolved into a longitudinal component $C_l({\bf q})$
 and a transverse component $C_t({\bf q})$ 
[see Eq.~(\ref{velocity})]; as a result, the self energy can be written 
as a sum of a contribution from longitudinal fluctuations and one
from transverse fluctuations, 
\begin{equation}
\hat{\Sigma}({\bf k}, \omega ) = \hat{\Sigma}_l({\bf k}, \omega ) 
             + \hat{\Sigma}_t({\bf k}, \omega ),
\end{equation}
where
\begin{equation}
\hat{\Sigma}_l({\bf k}, \omega ) =  \sum_{\bf q} 
   \left[m{\bf v}_{F}({\bf k})\cdot\hat{q}\right]^2 C_l({\bf q}) 
   \hat{ G}({\bf k-q}, \omega ),
\label{sigma_l}
\end{equation}
and
\begin{equation}
\hat{\Sigma}_t({\bf k}, \omega )= \sum_{\bf q}
   \left[m{\bf v}_{F}({\bf k})\times\hat{q}\right]^2 C_t({\bf q})
   \hat{ G}({\bf k-q}, \omega ). 
\label{sigma_t}
\end{equation}
The self-energy has both a momentum and frequency dependence.
Here we focus on the behavior of the self-energy near the Fermi
surface, assuming that it varies smoothly near the Fermi surface, and
we neglect the $\xi_{\bf k}$-dependence. Therefore the only 
momentum dependence is  through the angle $\phi$ on the
Fermi surface. 

\subsection{Longitudinal phase fluctuations}
\label{ss:long}

First we study the effect of the longitudinal phase fluctuations,
neglecting  the effect of vortices for now.  
Using Eqs.~(\ref{sigma_l}) and ({\ref{c_l}),
we obtain for the self-energy (see Appendix B for details)
\begin{eqnarray}
\hat{\Sigma}_l (\phi, \omega) 
&\approx &{4mT\over n_s(T)}\,{1\over 16\pi} \ln \left[
{\lambda ^2 + \tilde{\Delta}^2(\phi)+\tilde{\omega} ^2 \over 
\tilde{\Delta}^2(\phi)+\tilde{\omega} ^2} \right] \nonumber \\ 
&& \times \left( 
\begin{array}{cc} 
-i\tilde{\omega} & -\tilde{\Delta}(\phi) \\
-\tilde{\Delta}(\phi) & -i\tilde{\omega} \\
\end{array} \right),
\label{self1}
\end{eqnarray}
where $\tilde{\omega}$ and $\tilde{\Delta}$ can be calculated 
self-consistently by
\begin{equation}
\left( 
\begin{array}{cc} 
i\tilde{\omega} & -\tilde{\Delta}(\phi) \\
-\tilde{\Delta}(\phi) & i\tilde{\omega}\\
 \end{array} 
\right)
=\left( 
\begin{array}{cc} 
i\omega & -\Delta(\phi) \\
-\Delta(\phi) & i\omega \\
\end{array} 
\right)
-\hat{\Sigma}(\phi,  \omega).
\label{selfc}
\end{equation}
By analytically continuing the frequency $i\omega \rightarrow
\omega +i\eta $ in Eq. (\ref{selfc}),
we can calculate the single-particle
density of states
\begin{equation}
N(\omega) = -{1\over \pi}{\rm Im} \int {d\phi \over 2\pi}\int
d\xi_{\bf k} \,{\rm Tr}\, \hat{G}({\bf k},\omega).
\label{dos1}
\end{equation}

In order to apply this result to the underdoped cuprates, we need
to know the temperature dependence of the superfluid density $n_s(T)$
and the gap $\Delta_0(T)$. We neglect the temperature-dependence of
$\Delta_0(T) $, since the gap smoothly evolves into the pseudogap
above $T_c$.  A reasonable
form for the superfluid density that approximates the BCS theory is 
$n_s(T)/4m \approx n_s(0)/4m -\alpha \, T $ where
$\alpha$ is order of unity.\cite{Lee} 
At low temperatures, since the
reduction of the superfluid density is approximately linear in temperature,
this form is reasonable especially for underdoped cuprates
where $T_c$ is relatively low.
Throughout this paper we take $\alpha = 3/4$, but the exact value of
$\alpha$ is not crucial. Generally the value of $\alpha$ is dependent on
the value of $\Delta_0$.

The curves in Fig.~\ref{f:DOS}
show that the peak in the density of states (DOS) near $\omega = \Delta_0$ is
mildly smeared due to the phase fluctuations and the effect is more pronounced
at higher temperatures when the superfluid density is lower.
Notice that the DOS near $\omega =0$ is hardly affected even as the
temperature is raised, 
and therefore we expect that the electronic entropy is only
slightly increased due to the longitudinal phase fluctuation as illustrated
in Sec.~\ref{s:sph}.
\begin{figure}[t]
\epsfxsize=8.0cm \epsfbox{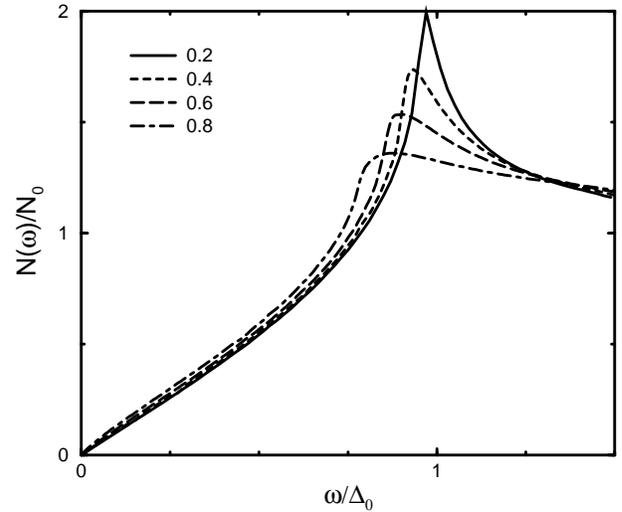}
\bigskip
\caption{Density of states in the presence of the longitudinal phase 
fluctuations at four different temperatures. 
Here $t = 4mT/n_s(0)$. Above $t=t_{BKT}\approx 0.67$,
the vortices are expected to be more important. Here we take $\lambda
=\Delta_0/v_F $. }
\label{f:DOS}
\end{figure}

\subsection{Transverse phase fluctuations}
\label{trans}

Since the two-dimensional superconducting transition is driven by 
vortex-antivortex pair unbinding, we need to take into account the effect 
of vortices,
especially above the vortex pair unbinding temperature $T_{BKT}$.\cite{bkt}
We assume that the longitudinal phase fluctuation is less important
in this temperature regime and consider only the effect of the
vortex distribution or the transverse superfluid velocity (see Appendix A). 
Using the Dyson equation similar to Sec.~\ref{ss:long} and neglecting the
$\xi_{\bf k}$-dependence in the self-energy,
from Eqs.~(\ref{sigma_t}) and (\ref{c_t}) we obtain the following form of the 
fermion self-energy
\begin{eqnarray}
\hat{\Sigma}_t (\phi,\omega) 
&\approx &-{\pi ^2 v_F^2\over 4\pi^2 K(l^* )\,
\xi^2_c\, e^{2l^*} }\int {d^2 q \over (2\pi )^2} \nonumber \\
&&{q_{\perp}^2 \over q^2 }\,{1\over q^2 +e^{-2l^* }/ \xi^2_c}\,
{1\over q_{\parallel}^2 v_F^2 + \tilde{\omega }^2+ \tilde{\Delta}^2(\phi)}
 \nonumber \\&&\times \left( 
\begin{array}{cc} 
i\tilde{\omega} & \tilde{\Delta}(\phi) \\
\tilde{\Delta}(\phi) & i\tilde{\omega} \\
\end{array} \right),
\end{eqnarray}
where $q^2=q_{\perp}^2+q_{\parallel}^2$.
Again using Eq.~(\ref{selfc}) we can self-consistently calculate 
$\tilde{\omega}$ and $\tilde{\Delta}$, and subsequently, the Green's function  
when $T>T_{BKT}$ by analytically continuing the frequency
$i\omega \rightarrow \omega +i\eta $.

The ARPES studies of  the cuprates show that the quasiparticle 
spectral function broadens dramatically when passing
from the superconducting to the normal state.\cite{arpes} How does 
vortex unbinding contribute to this broadening?
In Fig.~\ref{f:rate} we show the scattering rate
due to the vortices,
$1/\tau = {\rm Im} \, \tilde{\omega}$ at $\omega =  \Delta_0$ in the
three  directions ($\phi=0, \pi/8, \pi/12 $). 
   As the temperature is
raised from $T_{BKT}$,  $1/\tau $ rapidly increases to a value of order
$\Delta_0$, which is a rather large quantity. This shows that the
phase fluctuation is an important ingredient in the broadening of
the spectral function. As the temperature is further raised,
 we expect that the fluctuations in the gap magnitude $\Delta $
will further enhance the spectral width, which is beyond the
scope of discussion in this paper. 
\begin{figure}[t]
\epsfxsize=8.0cm \epsfbox{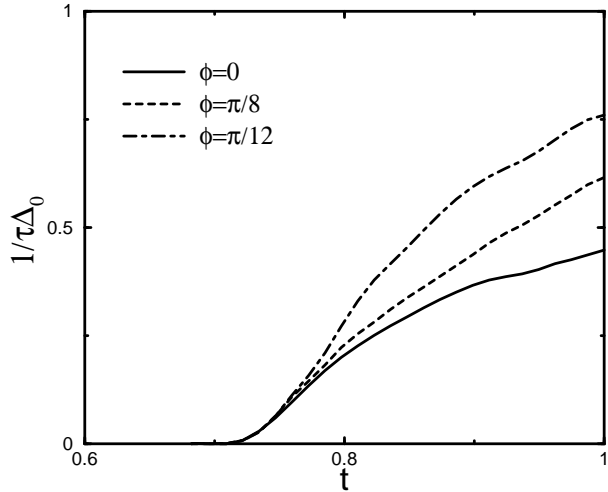}
\bigskip
\caption{The scattering rate $({\rm Im}\, \tilde{\omega})$ as a function of
$t = 4mT/n_s(0)$ at 
$\omega = \Delta_0 $ and in the
three directions $\phi_1=0,\, \phi_2=\pi/8,\, \phi_3=\pi/12$, 
in the presence of unbound vortex-antivortex pairs.}
\label{f:rate}
\end{figure}
So far we have discussed the spectral properties of the {\it neutral} fermions.
Of much interest is the spectral broadening of the physical electrons
as revealed by the ARPES data.  
For instance, in the absence of the interaction 
term $S_I$ in the action, the Green's function for the physical electrons
factors as 
\begin{eqnarray}
\langle c_\sigma({\bf x},\tau) c^\dagger_\sigma ({\bf 0},0)\rangle 
&=& \langle e^{i\theta ({\bf x},\tau)/2} e^{-i\theta ({\bf 0},0)/2}\rangle
\nonumber \\ &&\times  
\langle \psi_\sigma ({\bf x},\tau) \psi^\dagger_\sigma ({\bf 0},0)\rangle.
\label{decouple}
\end{eqnarray}
We may gain insight  into the
form of the electron spectral function from Eq.~(\ref{decouple}).
Above $T_{BKT}$, the correlation function
$\langle e^{i\theta({\bf x})/2}\, e^{-i\theta({\bf 0})/2} \rangle $
behaves approximately as $e^{-|{\bf x}|/4\xi_+}$, where $\xi_+(T) $ is the
BKT correlation length,\cite{corrl}
and the electron Green's function is roughly
\begin{equation}
\langle c({\bf x})\, c^{\dag}({\bf 0}) \rangle \sim
e^{-|{\bf x}|/4\xi_+}\, {\cal G}({\bf x}) .
\label{factors}
\end{equation}
In the low-energy effective theory we consider here, we may think of
the quasiparticles as following a quasiclassical trajectory with a
Fermi momentum ${\bf k}_F$. 
Therefore, the leading contribution to the two-point
correlation function $\langle c({\bf x})\, c^{\dag}({\bf 0}) \rangle $
comes from the particles with the Fermi momentum ${\bf k}_F$ in parallel
with the vector $\pm {\bf x}$. Then we may treat ${\bf x}$ and
${\bf k}_F$ as effectively one-dimensional.\cite{KHM}
The one-dimensional Fourier transform of Eq.~(\ref{factors})
shows that the spectral function of the electrons is additionally 
broadened by the width $ v_F/4\xi_+ $---the electronic 
spectral width is roughly  the sum of $ v_F/4\xi_+ $ and the spectral width of 
the neutral fermions.
\begin{figure}[t]
\epsfxsize=8.0cm \epsfbox{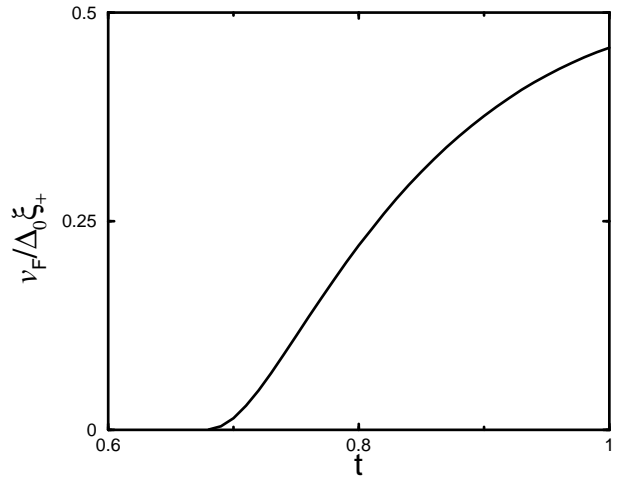}
\bigskip
\caption{Inverse of BKT correlation length vs reduced temperature.
}
\label{f:corrl}
\end{figure}

As a result of the rapid spectral broadening above $T_{BKT}$, the 
single-particle DOS               is significantly modified and
the singularity at $\omega =\Delta_0$ in the DOS is weakened and broadened.
Unlike the case of spectral functions, the neutral fermion DOS
coincides with that of physical electrons.
Fig.~\ref{f:vdos} shows the DOS in the maximum gap direction at four
different temperatures. When $t > 0.67$, the DOS is rapidly smeared
out and the gap is filling in, exhibiting a pseudogap-like behavior.
\begin{figure}[ht]
\epsfxsize=8.0cm \epsfbox{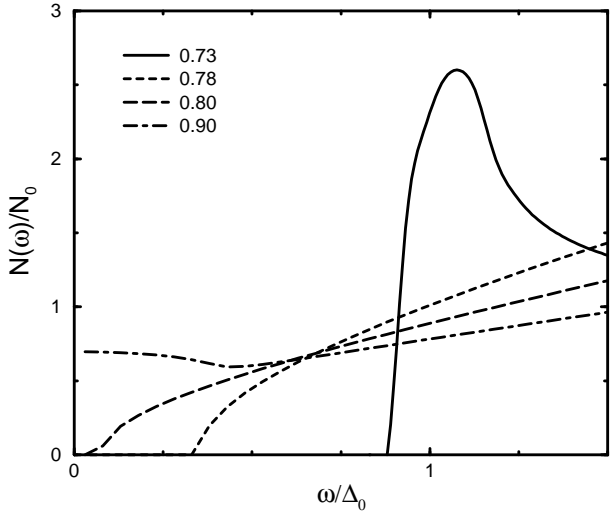}
\bigskip
\caption{Density of states at $\phi =0$ (in the maximum gap direction)
at four temperatures above $t_{BKT}=0.67$
when the unbound vortex-antivortex pairs are taken into account.}
\label{f:vdos}
\end{figure}

\section{Thermodynamic properties}
\label{s:sph}
Experiments on $\rm Y_{0.8}Ca_{0.2}Ba_{2}Cu_3 O_{7-\delta}$
show that there is a BCS-like discontinuity in the specific heat at optimal 
doping.\cite{specific-heat}
However, in the underdoped regime, the superconducting transitions
are marked by weaker and broader peaks in the specific heat. 
With this puzzling feature in mind, we consider the effect of the phase
fluctuations on specific heat in this section.

Before we proceed,
we consider one consequence of the pseudogap phenomenon in underdoped
cuprates.  We can estimate the specific heat change in the presence of
the pseudo gap
using the BCS mean-field theory;\cite{AGD} the 
superconducting-normal free energy difference is  
\begin{equation}
\Delta F = F_S-F_N = \int_0^{\Delta_s(T)} {d(1/|g|)\over d\Delta_s }
\Delta_s ^2 \, d\Delta_s,
\label{FreeE}\end{equation}
where
\begin{equation}
{1\over |g|} = {mk_F\over 2\pi^2} \int_{0}^{\omega_D} d\xi\, 
{\tanh(\sqrt{ \xi^2+ \Delta^2(T)}/2T  ) \over \sqrt{ \xi^2+ \Delta^2(T)}}. 
\label{gapeq} 
\end{equation}
Here $g $ is the BCS pair coupling and $\Delta_s$ and $\Delta $
are the superconducting gap and the total electronic gap respectively.
The ARPES\cite{arpes} and STS\cite{sts} data show that
the electronic gap evolves smoothly through $T_c$, from the superconducting
gap to the pseudogap. Since $\Delta(T)$ does not rapidly change at $T_c$
and Eq.~(\ref{gapeq}) is expressed in terms of $\Delta(T)$ only, 
$d(1/|g |)/d\Delta_s $ is a very small number and there is
no dramatic increase in entropy at $T_c$. Therefore, we do not
expect a BCS-like specific heat jump.

From the above discussion we see that most of the characteristic changes in the
 specific heat 
at $T_c$ should come from the order
parameter fluctuations. Here we consider the
effect of the phase fluctuations at both $T > T_{BKT}$ and $T < T_{BKT}$. 
First we  consider the
electronic entropy change due to the longitudinal phase fluctuations
which is more important at $T< T_{BKT}$. From the
Green's function and the density of states
 we obtained in Eq.~(\ref{self1}) and Eq. (\ref{dos1}), we may
approximately express the entropy as
\begin{eqnarray}
S&\approx & -\int_0^\infty d\omega\, N(\omega)\left\{f(\omega)  \ln f(\omega) \right.
\nonumber \\ &&\left.  +
[1-f(\omega)] \ln [1-f(\omega)] \right\},
\end{eqnarray}
where $N(\omega)$ is the single-particle density of states and
$f(x) = 1/(e^{\beta x}+1)$.
We can compare the entropy thus obtained with the mean-field BCS
entropy, as illustrated in Fig.~\ref{f:ent1}. 
We observe that the electronic entropy change is so minute
that it may be hardly visible in the specific heat. The
small entropy increase is due to the fact that the density of
states is affected by the phase fluctuations only when $\omega \approx 
\Delta_0$ (see Fig.~\ref{f:DOS});
 the electronic entropy is  insensitive to the longitudinal phase fluctuations.  
\begin{figure}[t]
\epsfxsize=8.0cm \epsfbox{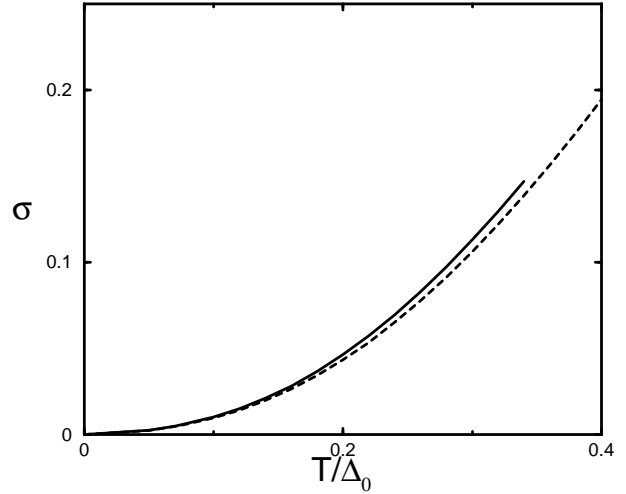}
\bigskip
\caption{The reduced electronic
 entropy $\sigma = ({2\pi E_F\over \Delta_0 k_B })\, S$
vs the temperature. $\sigma $ has the meaning of the entropy in an
area of $1/k_F^2 $ normalized by the maximum gap energy $\Delta _0$.
The dotted line is the electronic entropy obtained using the $d$-wave
BCS theory, and the solid line is the entropy in the presence of the
longitudinal phase fluctuations. Here the case of $n_s(0)/4m = 
\Delta_0/3 $ is illustrated.
 }
\label{f:ent1}
\end{figure}

Now we turn to the behavior of specific heat in the
presence of unbound vortex-antivortex pairs
at $T>T_{BKT}$.  In the 2D XY model, the vortex-antivortex
dissociations cause an unobservable singularity at the transition
temperature, followed by a broad bump at a temperature higher than
$T_{BKT}$.\cite{Berker}
In conventional superconductors, the bump in the  specific heat
due to the vortex unbinding may be unobservable---the
density of dissociated vortices can be only so large before they
saturate the whole sample
since the vortex core size
is much larger than the electronic mean spacing. 
However, in the cuprate superconductors,
the vortex core size is only a few times the lattice spacing, so
the vortex dissociation may give a sizable contribution to the
specific heat. To reduce the theory down to the effective vortex 
degrees of freedom,
we integrate out the fermion fields and express the action in the form 
in Eq.~(\ref{Sv}). Then we can evaluate the specific heat by
\begin{equation}
C = k_B~{d\over dT}\Big{(} T^2 {d\over dT} \bar{\cal F} \Big{)}
\end{equation}
where the reduced free energy is defined as
\begin{equation}
\bar{\cal F} = \ln {\cal Z} =\ln \Big{(} \int {\cal D} n \, e^{-S_{\rm v}}
\Big{)}.
\end{equation}
The free energy is a function of the temperature through
$K(0)$ and $y(0)$ as defined in Appendix A. We evaluate $\bar{\cal F}$
following the prescription by Berker and Nelson,\cite{Berker} who
computed the specific heat in the 2D XY model. The only difference here is 
that in our model the superfluid density and the vortex core energy
have a temperature dependence whereas in the spin XY model the
spin stiffness $JT$ is a constant.
\begin{figure}[t]
\epsfxsize=8.0cm \epsfbox{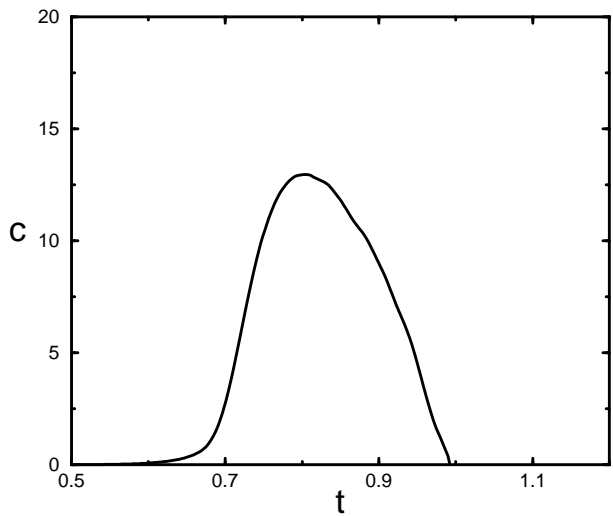}
\bigskip
\caption{The reduced heat capacity per area of a vortex core 
$( c= (\xi^2_c /k_B)~C )$ vs
temperature $(t = 4mT/n_s(0) )$. There is a very weak singularity at 
$t=0.67$.}
\label{f:heat}
\end{figure}

The result is shown in Fig.~\ref{f:heat}. The heat capacity per
vortex core area $\xi ^2_c$ is peaked at $t \approx 0.8$ and the peak
height about $13k_B $. Now we may cast this result on the electronic
length scale. Since $k_F \xi_c \approx 10$ in cuprates, the heat 
capacity per unit cell of area $1/k_F^2$ is about $0.1 k_B$, which
is a large number compared to the electronic specific heat.
In conventional low-$T_c$ superconductors, there is a factor of 
$10^{-6}$ reduction in the peak since $ k_F \xi_c {> \atop \sim}10^3 $. 
The specific heat peak
may be practically unobservable in this case.
We conclude that if there is a vortex pair unbinding transition
in underdoped cuprates, its effect on thermodynamic quantities
would be visible, although the actual specific heat measurement may be
complicated by other sources of thermal fluctuations.

\section{Conclusions}

The low-energy effective theory which we have applied to the underdoped
cuprates is a convenient model  which describes
a complex many-body system (whose microscopic details are unknown)
in terms of a few macroscopic parameters such as the superfluid density,
effective mass, and the magnitude of the superconducting gap.
Utilizing this phenomenological theory,
we are able to calculate the spectral properties in the presence 
of  phase fluctuations without regard to the microscopic details.
By assuming a robust quasiparticle gap and
a small superfluid density, as appropriate for the underdoped cuprates, 
we found that the vortex pair unbinding
transition causes pronounced effects on the single-particle properties:
(i) The peak in the spectral function is dramatically broadened as the
temperature  is raised through $T_{BKT}$; (ii) The density of states
is widely smeared out at $T>T_{BKT}$, significantly deviating from
the usual superconducting state although the gap magnitude is robust.
These two features are reminiscent of the pseudogap behavior in
underdoped cuprates, as observed in the ARPES\cite{arpes} and STS\cite{sts}
experiments. It is remarkable that we can reproduce a few of the pseudogap-like
properties by including only the phase degree of freedom.
This indicates that along with the fluctuations in the order parameter
amplitude, the phase fluctuations may be an essential ingredient in the
pseudogap phenomena. For instance, phase fluctuations may better
explain the fact that the superconducting gap nodes evolve into
extended gapless Fermi arcs in the pseudogap phase since the phase
fluctuations are more significant  near the nodes 
(Fig. \ref{f:rate}). The amplitude fluctuations alone may not be
sufficient to explain this.\cite{later} 

The role  of phase fluctuations in the pseudogap regime has recently been explored in several 
approaches which bear some similarity to ours.  As mentioned in Sec.~II, Franz and 
Millis\cite{franz98} have also studied the effect of classical phase fluctuations on the 
spectroscopic properties of the underdoped cuprates,  by coupling the 
$d$-wave quasiparticles to the fluctuating supercurrents due to unbound
vortex-antivortex pairs.  While our approach to treating this coupling is 
quite different from the approach  of Franz and Millis,  our conclusions 
are similar---that transverse phase fluctuations are important in determining the
spectral properties above $T_c$.  
On a different tack, Balents, Fisher, and 
Nayak\cite{balents98} have recently proposed a zero-temperature ``nodal liquid theory''
in which there is a  transition between the superconducting phase and the 
nodal liquid (tentatively identified as the ground state of the pseudogap regime)
which is driven by {\it quantum} fluctuations of the 
order-parameter phase.  The properties of this model at non-zero temperatures
have yet to be investigated. 

Before closing we should note that our interpretation of the pseudogap regime 
as the remnant of a superconducting phase which has been destroyed by phase
fluctuations is not without criticism. 
Geshkenbein  {\it et al.}\cite{geshkenbein97} 
and Randeria\cite{randeria97} have argued that the fluctuation effects should be
significant over a much wider temperature range than is observed experimentally,
and that strong pairing correlations (beyond BCS theory) need to be incorporated 
into any model of the pseudogap regime.  A completely different explanation of the 
pseudogap behavior has been proposed by Lee and Wen,\cite{Lee} who argue 
that the superconducting phase
is destroyed by the thermal excitation of nodal quasiparticles, which quenches the 
superfluid density while retaining the quasiparticle gap.  Finally, we have 
neglected the Coulomb interaction in  our model; including it would 
presumably suppress the longitudinal
phase fluctuations while leaving the transverse fluctuations 
unaffected.\cite{franz98}  All of these criticisms merit further investigation
in a more detailed study of the pseudogap phenomenon.

\section*{Acknowledgments}

We would like to thank
Mohit Randeria for many helpful discussions on the phenomenology of the
underdoped cuprates; 
Marcel Franz and Wilhelm Zwerger for their helpful comments on the manuscript;
and Matthew Fisher for pointing out possible difficulties with the gauge 
transformation in the presence of vortices.
This work was supported by NSF grant DMR 96-28926 and by the 
National High Magnetic Field Laboratory.  

\appendix

\section{Longitudinal and transverse fluctuations of the superfluid velocity}

For the sake of completeness, in this Appendix we have collected together 
some of the
important results regarding the Berezinskii-Kosterlitz-Thouless (BKT) transition
\cite{bkt} which pertain to our calculation of the fermion self-energy.
A more detailed discussion can be found in Ref.~\onlinecite{chaikin}.
First, we resolve the phase field into longitudinal (spin wave) and transverse 
(vortex) components.
This is  done by writing $\theta = \theta_a + \theta_v$, with $\theta_a$
the analytic spin-wave contribution and $\theta_v$ the singular vortex 
contribution. The superfluid velocity can then be decomposed as
${\bf v}_s=(1/2m)\nabla \theta = {\bf v}_s^{l} + {\bf v}_s^{t}$, 
with ${\bf v}_s^{l} = (1/2m)\nabla \theta_a$ and
${\bf v}_s^{t} = (1/2m)\nabla \theta_v$, such that 
$\nabla\times {\bf v}_s^{l}=0$ and $\nabla\cdot {\bf v}_s^{t}=0$.
The classical phase action then becomes
\begin{eqnarray}
S &= &{m n_s(T)\over 2T}\int d^2x \, v_s^2 \nonumber \\ 
  &=& {m n_s(T)\over 2T}\int d^2x \, [(v_s^l)^2 + (v_s^t)^2],
\label{action}
\end{eqnarray}
since the cross term is  zero as shown below.

The curl of the transverse component is the total vorticity; for
singly quantized vortices with circulation $\pm 2\pi\hbar/2m$, this
can be written as
\begin{equation}
\nabla\times {\bf v}_s^{t} = {2 \pi \over 2 m} n({\bf x}) {\hat z},
\end{equation}
where the vortex density $n({\bf x})$ is 
\begin{equation}
n({\bf x})=\sum_i q_i \delta({\bf x}-{\bf x}_i),
\end{equation}
with vortices at positions $\{ {\bf x}_i\}$ and with 
vortex ``charges'' $q_i=\pm 1$ (here we consider only neutral 
vortex configurations such that $\sum_i q_i = 0$). Using the 
fact that ${\bf v}_s^t$ is transverse, 
\begin{equation}
{\bf v}_s^{t}({\bf x}) = \nabla \times \int d^2x'\, G({\bf x}-{\bf x}') 
{2 \pi \over 2 m} n({\bf x'}) {\hat z},
\end{equation}
where $\nabla^2 G = - \delta({\bf x} - {\bf x}')$; Fourier transforming, 
\begin{equation}
{\bf v}_s^{t}({\bf q}) = i {{\bf q}\times \hat{z} \over q^2} 
{2\pi\over 2m} n({\bf q}).
\label{transverse}
\end{equation}
The analogous expression for the longitudinal component is
\begin{equation}
{\bf v}_s^{l}({\bf q}) = {i{\bf q} \over 2m} \theta_a({\bf q}). 
\end{equation} 
It is easy to verify that the cross term 
${\bf v}_s^{t}({\bf q})\cdot {\bf v}_s^{l}({\bf q})$
in Eq. (\ref{action}) is zero.
The final expression for the vortex contribution to the classical phase action 
is obtained by substituting Eq.~(\ref{transverse})
into Eq.~(\ref{action}):  
\begin{eqnarray}
S_{\rm v} &=& \pi K(0) \mathop{\int \int}\limits_{|{\bf x}-{\bf x}'|>\xi_c} 
d^2x\, d^2x'\, 
n({\bf x}) n({\bf x^{\prime}})\ln (|{\bf x-x^{\prime}}|/ \xi_c) \nonumber \\
&&\quad +\ln y(0) \int d^2x\, n^2({\bf x})\nonumber \\
&= & {1\over 2 } \sum_{\bf q} \left\{ {4\pi ^2 K(0) \over  q^2}
-8\pi \xi^2_c \ln \left[ y(0) \right] \right\}n({\bf q})n({\bf -q}),
\label{Sv}
\end{eqnarray}
where $K(0) = n_s(T)/4mT$ is the bare spin stiffness, 
$y(0) = \exp(-{\cal E}_c/4\pi T)$ is the bare fugacity
with ${\cal E}_c = T\, K(0)\ln \kappa $ the vortex core energy,
$\xi_c \sim v_F/\Delta_0$ is the vortex core size, and
$\kappa $ is the ratio of the magnetic penetration depth to the
coherence length.
\begin{figure}[t]
\epsfxsize=8.0cm \epsfbox{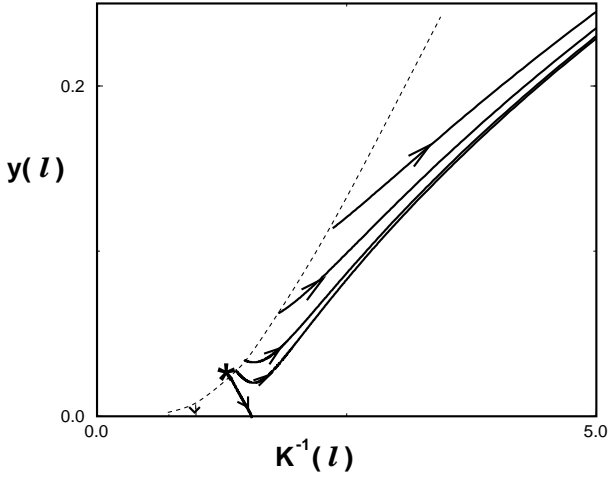}
\bigskip
\caption{The RG flow of $(K^{-1}(l), y(l) )$. The dashed curve 
is the locus of initial values $(K^{-1}(0), y(0) )$. The asterisk (*)
marks the vortex pair unbinding transition temperature ($T_{BKT}$) 
which in this model corresponds
to $4mT_{BKT}/n_s(0)\approx 0.67$. Above this transition temperature, 
$(K^{-1}(0), y(0) )$ rides on the outgoing curves and $(K^{-1}(l), y(l) )$
values diverge in the $l \rightarrow \infty $ limit.}
\label{f:flow}
\end{figure}

In this model there is a critical temperature $T_{BKT}$ 
above which the two-dimensional
quasi-long-range order is destroyed by thermal fluctuations.
At $T<T_{BKT}$,  the vortices are bound into neutral pairs  
(when there is no external magnetic field), and the superfluid density
is non-zero. In this temperature regime, consideration of the
soft longitudinal phase fluctuation $\theta_a$ is sufficient.
When $T>T_{BKT}$, the bound pairs of vortices dissociate and 
free vortices proliferate as the temperature is raised, driving the superfluid
 density to zero.  This picture is made quantitative using the
RG flow equations \cite{bkt} for the coupling constants
$(K(l),y(l))$ at the length scale $e^l$ (see Fig. \ref{f:flow}):
\begin{eqnarray}
{dK^{-1}(l) \over dl} &=& 4\pi ^3 y^2(l), \nonumber \\
{dy(l) \over dl} &=& \big{[} 2-\pi K(l) \big{]} y(l).
\label{rg}
\end{eqnarray}

To calculate the fermion self-energy we need the vortex-vortex
correlation function 
$\langle n({\bf q})n({\bf -q}) \rangle \equiv \Gamma_2({\bf q},K(0),y(0))$,
which can be calculated approximately using the 
Debye-H\"uckel theory developed by Halperin and Nelson.\cite{Halperin}
To implement this, we first note that under scale transformations the 
correlation function scales as 
\begin{equation}
\Gamma_2 \big{(}{\bf q}, K(0),y(0 ) \big{)} = e^{-2l}\, 
\Gamma_2 \big{(}e^l {\bf q}, K(l ),y(l) \big{)}.
\end{equation}
Now we integrate the RG flow equations out to a length scale
$l^*$ where the density of free vortices is high; at this scale 
we may integrate over the coarse-grained vortex density rather 
than the individual vortex coordinates 
(the Debye-H\"uckel approximation).  Since the 
classical phase action, Eq.~(\ref{Sv}), is quadratic in the vortex density, 
the correlation function is 
\begin{eqnarray}
\Gamma_2 \left( {\bf q}, K(0),y(0 ) \right) 
= {e^{-2l^*}\over 4\pi^2 K(l^*)/(e^{l^*}q)^2 - 8\pi \xi^2_c \ln y(l^*) }. 
\end{eqnarray}
If we choose $l^* $ such that 
\begin{equation}
4\pi ^2 K(l^* )  \xi^2_c = -8\pi \xi^2_c \ln y(l^*),
\end{equation}
then the correlation function takes the particularly simple
form
\begin{equation}
\Gamma_2 \left( {\bf q}, K(0),y(0 ) \right) =
{1\over 4\pi^2 K(l^* )}\, {1\over 1/q^2 +\xi^2_c e^{2l^*}}.
\end{equation}
Both $l^*$ and $K(l^*)$ are determined from a numerical integration of the 
RG equations, Eq.~(\ref{rg}).

Using these results we can calculate the velocity-velocity correlation 
function for $T>T_{BKT}$, 
\begin{eqnarray}
\langle v_s^{\alpha}({\bf q}) v_s^{\beta}(-{\bf q})\rangle 
  &=& C_l({\bf q})\, \hat{q}_\alpha\hat{q}_\beta  \\ \nonumber
 & & \quad   + C_t({\bf q})\, (\delta_{\alpha\beta} - \hat{q}_\alpha\hat{q}_\beta),
\label{velocity}
\end{eqnarray}
where the longitudinal component is 
\begin{eqnarray}
 C_l({\bf q})&=&{q^2 \over 4 m} \langle \theta_a({\bf q})
 \theta_a(-{\bf q})\rangle
 \\ \nonumber 
 & = & { T\over 4 m^2 n_s(T)},
\label{c_l}
\end{eqnarray}
and the transverse component is 
\begin{eqnarray}
C_t({\bf q})&=&{\pi^2 \over m^2 } {\langle n({\bf q}) n(-{\bf q})\rangle \over q^2}
 \\ \nonumber 
 & = & {1\over 4 m^2 K(l^*)} {1\over 1 + \xi_c^2 e^{2l^*} q^2}. 
\label{c_t}
\end{eqnarray}

\section{Details of the calculation of the self-energy}

In this Appendix we provide some of the details of the 
calculation of the self energy in Eqs.~(\ref{self1}) and (\ref{selfc}).

Near the Fermi surface, neglecting the $\xi_{\bf k}$-dependence in
$\Sigma ({\bf k},\omega )$,
\begin{eqnarray}
\hat{G}^{-1}({\bf k}, \omega) 
&=& \left( 
\begin{array}{cc} 
i\tilde{\omega}-\xi_{\bf k} & -\tilde{\Delta}(\phi) \\
-\tilde{\Delta}(\phi) & i\tilde{\omega}+\xi_{\bf k} \\
\end{array} \right) \nonumber \\
&=& \left( 
\begin{array}{cc} 
i{\omega}-\xi_{\bf k} & -{\Delta}(\phi) \\
-{\Delta}(\phi) & i{\omega}+\xi_{\bf k} \\
\end{array} \right)-\hat{\Sigma}(\phi , \omega ) ,
\end{eqnarray}
where $\phi$ is the angle on the Fermi surface. From Eq. (\ref{dyson}),
when $|{\bf k}|=k_F$,
\begin{eqnarray}
\hat{\Sigma}_l(\phi , \omega ) &\approx &-
\sum_{\bf q} \Big{[}m { {\bf v}_F({\bf k})\cdot \hat{ q} } \Big{]} ^2
C_l({\bf q})\left( \begin{array}{cc} 
 i\tilde{\omega}+\delta\xi_{\bf q} & \tilde{\Delta}(\phi) \\
\tilde{\Delta}(\phi) & i\tilde{\omega}-\delta\xi_{\bf q} \\
\end{array} \right) \nonumber \\
&&\times {1\over \tilde{\omega }^2 + 
(\delta\xi_{\bf q})^2 + \tilde{\Delta}^2(\phi) } ,
\end{eqnarray}
where $\delta\xi_{\bf q} = {\bf v}_F({\bf k})\cdot {\bf q}$.
Now, using the integral
\begin{eqnarray}
&&\int dx\, dy\,  {x^2 \over x^2+y^2}\, {1\over x^2+a^2} 
\nonumber \\ &\approx&
\int _{-\lambda}^{\lambda}dx \int _{-\infty}^{\infty}dy\, 
{x^2 \over x^2+y^2}\, {1\over x^2+a^2}
\nonumber \\&=& \pi\ln \Big{(} {\lambda ^2+a^2 \over
a^2 }\Big{)} ,
\end{eqnarray}
we obtain Eq.~(\ref{self1}).


\begin{references}
\bibitem{mpaf} In has been pointed out to us (M.P.A. Fisher, private
communication) that in the presence of unbound, singly quantized vortices 
this gauge transformation is singular, and therefore multi-valued. 
One can in principle remedy this difficulty by defining branch cuts in the 
two dimensional plane which 
terminate on the vortices; however, in practice it is unclear how one would 
incorporate this cut structure into a perturbative many-body calculation,
such as the one which we develop in this work. 
We leave this as an unresolved issue which deserves further study.
\end{references}
\end{document}